\documentclass[prl,twocolumn,fleqn]{revtex4-1}
\def\pr{1} \def\letter{1}

\usepackage{epsfig}
\usepackage{epstopdf}
\usepackage{graphicx}
\usepackage{ifthen}

\usepackage{amsmath}
\usepackage[psamsfonts]{amssymb}
\usepackage{euscript}

\usepackage{latexsym}

\newskip\humongous \humongous=0pt plus 1000pt minus 1000pt

\newif\ifdtup

\def\hsp{,\hspace{.7cm}}

\ifthenelse{\equal{\pr}{1}}{

\renewcommand{\cos}{\textrm{cos}}
\renewcommand{\sin}{\textrm{sin}}

\renewcommand{\(}{\begin{equation}}
\renewcommand{\)}{end{equation} \vspace{-.05in}\linebreak}

\newcounter{saveeqn}
\newcounter{savealpheqn}

\newcommand{\alpheqn}{\setcounter{saveeqn}{\value{equation}}%
  \stepcounter{saveeqn}\setcounter{equation}{0}%
  \renewcommand{\theequation}{\mbox{\arabic{section}.\arabic{saveeqn}
\alph{equation}}}
  \renewcommand{\)}{\end{equation}}}
\def\part#1{\frac{\partial}{\partial{#1}}}%
\def\group#1{\refstepcounter{equation}\setcounter{saveeqn}
 {\value{equation}}%
  \label{#1}\setcounter{equation}{0}%
\renewcommand{\theequation}{\mbox{\arabic{section}.\arabic{saveeqn}
\alph{equation}}}
  \renewcommand{\)}{\end{equation}}}
\newcommand{\reseteqn}{\setcounter{equation}{\value{saveeqn}}%
  \renewcommand{\theequation}{\arabic{section}.\arabic{equation}}%
  \renewcommand{\)}{\end{equation}}}

\newcommand{\aalpheqn}{\setcounter{saveeqn}{\value{equation}}%
  \stepcounter{saveeqn}\setcounter{equation}{0}%
  \renewcommand{\theequation}{\mbox{
        \Alph{subsection}.\arabic{saveeqn}\alph{equation}}}
   \renewcommand{\)}{\end{equation}}}
\newcommand{\areseteqn}{\setcounter{equation}{\value{saveeqn}}%
  \renewcommand{\theequation}{\Alph{subsection}.\arabic{equation}}%
  \renewcommand{\)}{\end{equation}}}

\renewcommand{\thefootnote}{\alph{footnote}}
\renewcommand{\(}{\begin{equation}}
\renewcommand{\)}{\end{equation}}
\newcommand{\ba}{\begin{eqnarray}}
\newcommand{\ea}{\end{eqnarray}}

\newcommand{\bp}{\mathop{\vtop{\ialign{##\crcr
   $\hfil\displaystyle{}\hfil$\crcr\noalign{\kern-13pt\nointerlineskip}
   \BIG{(}\hskip0pt\crcr\noalign{\kern3pt}}}}}
\newcommand{\cbp}{\mathop{\vtop{\ialign{##\crcr
   $\hfil\displaystyle{}\hfil$\crcr\noalign{\kern-13pt\nointerlineskip}
   \BIG{)}\hskip0pt\crcr\noalign{\kern3pt}}}}}
\newcommand{\pa}{\mathop{\vtop{\ialign{##\crcr
    
$\hfil\displaystyle{\oplus}\hfil$\crcr\noalign{\kern+1pt\nointerlineskip 
}
   \hspace{.08in}$^{\alpha=0}$\hskip6pt\crcr\noalign{\kern3pt}}}}}
\renewcommand{\hsp}{,\hspace{.3in}}

\numberwithin{equation}{section}
\renewcommand{\theequation}{\mbox{\arabic{equation}}}

\catcode`\@=11
\def\vereq#1#2{\lower3pt\vbox{\baselineskip1.5pt \lineskip1.5pt
\ialign{$\m@th#1\hfill##\hfil$\crcr#2\crcr\sim\crcr}}}
\catcode`\@=12

\renewcommand{\(}{\begin{equation}}
\renewcommand{\)}{\end{equation}}

\newcommand{\beas}{\begin{eqnarray*}}
\newcommand{\eeas}{\end{eqnarray*}}

\newcommand{\bquo}{\begin{quote}}
\newcommand{\enqu}{\end{quote}}

}{}

\newcommand{\C}{{\mathbb C}}
\newcommand{\cp}{{\mathrm{\mathbb CP}}}

\newcommand{\beq}{\begin{equation}}
\newcommand{\eeq}{\end{equation}}
\newcommand{\bea}{\begin{eqnarray}}
\newcommand{\eea}{\end{eqnarray}}

\newskip\humongous \humongous=0pt plus 1000pt minus 1000pt

\newif\ifdtup

\def\noprl#1{\ifthenelse{\equal{\pr}{1}}{}{#1} }

\ifthenelse{\equal{\letter}{0}}{ 

\catcode`\@=11

\@addtoreset{equation}{section}
\def\theequation{\thesection.\arabic{equation}}

\def\@normalsize{\@setsize\normalsize{15pt}\xiipt\@xiipt
\abovedisplayskip 14pt plus3pt minus3pt%
\belowdisplayskip \abovedisplayskip
\abovedisplayshortskip \z@ plus3pt%
\belowdisplayshortskip 7pt plus3.5pt minus0pt}

\def\small{\@setsize\small{13.6pt}\xipt\@xipt
\abovedisplayskip 13pt plus3pt minus3pt%
\belowdisplayskip \abovedisplayskip
\abovedisplayshortskip \z@ plus3pt%
\belowdisplayshortskip 7pt plus3.5pt minus0pt
\def\@listi{\parsep 4.5pt plus 2pt minus 1pt
      \itemsep \parsep
      \topsep 9pt plus 3pt minus 3pt}}

\relax

\catcode`@=12

\catcode`\@=11

\def\section{\@startsection{section}{1}{\z@}{3.5ex plus 1ex minus
    .2ex}{2.3ex plus .2ex}{\large\bf}}

\def\thesection{\arabic{section}}
\def\thesubsection{\arabic{section}.\arabic{subsection}}

\def\appendix{\setcounter{section}{0}
  \def\thesection{Appendix \Alph{section}}
  \def\thesubsection{\Alph{section}.\arabic{subsection}}
  \def\theequation{\Alph{section}.\arabic{equation}}}

}{} 

\begin{document}
\def\thefootnote{\fnsymbol{footnote}}
\def\thetitle{The Compactified Principal Chiral Model's Mass Gap}
\def\autone{Jarah Evslin}
\def\auttwo{Baiyang Zhang}
\def\affa{Institute of Modern Physics, NanChangLu 509, Lanzhou 730000, China}
\def\affb{University of the Chinese Academy of Sciences, YuQuanLu 19A, Beijing 100049, China}
\def\affc{Wigner Research Centre, H-1525 Budapest 114, P.O.B. 49, Hungary}

\ifthenelse{\equal{\pr}{1}}{
\title{\thetitle}
\author{\autone}
\affiliation {\affa}
\affiliation {\affb}
\author{\auttwo}
\affiliation {\affc}
}{
\begin{center}
{\large {\bf \thetitle}}

\bigskip

\bigskip

{\large \noindent  \autone{${}^{1,2}$} \footnote{jarah@impcas.ac.cn} and \auttwo{${}^3$} \footnote{baiyang.zhang@wigner.mta.hu } }

\vskip.7cm

1) \affa\\
2) \affb\\
3) \affc\\

\end{center}
}

\begin{abstract}
\noindent

\noindent
If the space of minima of the effective potential of a weakly coupled 2d quantum field theory is not connected, then a mass gap will be nonpertubatively generated.  As examples, we consider two $\sigma$ models compactified on a small circle with twisted boundary conditions. In the compactified $\cp^1$ model the vacuum manifold consists of two points and the mass gap is nonperturbative.  In the case of the compactified SU(2) principal chiral model the vacuum manifold is a single circle and the mass gap is perturbative.

\end{abstract}

%
\setcounter{footnote}{0}
\renewcommand{\thefootnote}{\arabic{footnote}}


\ifthenelse{\equal{\pr}{1}}{
\maketitle
}{}


The similarity between the topological structure of fractional instantons in the 2-dimension $\cp^1$ sigma model and in Yang-Mills theory has long led to speculations that they play [distinct] roles in the generation of the mass gaps of both theories  \cite{merons}.  Intriguingly a similar half-charged excitation appears to cause the mass gap in the SU(2) principal chiral model (PCM), where the Euclidean theory has no topologically stable solutions.  More precisely, the mass gap has been found analytically \cite{janosh} and on the lattice \cite{latgap} to be proportional to the strong coupling scale which is the exponential of half of the action of the uniton saddle point found in Ref.~\cite{uniton}.  Recently, in two remarkable papers~\cite{danielecorto,daniele} the authors have proposed a new window on this puzzle.  They claim that a weakly-coupled circle compactification of the PCM with certain boundary conditions is adiabatically connected to the original model.  

Needless to say, if crossed, the adiabatic bridge constructed by the authors may allow the mass gap of the PCM to be understood and perhaps to shed light on confinement in Yang-Mills.  As a first step in this direction, in the current note we will attempt to understand the weakly coupled (small circle) side of this bridge.  We find several surprises with respect to its expected properties.  We apply the same analysis to the $\cp^1$ model, whose adiabatic compactification was introduced in Refs.~\cite{cplungo,cpcorto}.  The Hamiltonian which we find for the resulting quantum mechanics is similar to but distinct from that found in Ref.~\cite{cplungo}.  This Hamiltonian provides a starting point for future investigations of the nonperturbative nature of the adiabatically compactified $\cp^1$ model.

The SU(2) principal chiral model is a $\sigma$ model whose target space is the group manifold SU(2).  Let $U$ be the SU(2)-valued field.  Consider the $\sigma$ model compactified on a circle of circumference $L$ with the adiabatic twisted boundary conditions of Refs.~\cite{danielecorto,daniele}
\beq
U\left(\frac{L}{2}\right)=\sigma_3 U\left(-\frac{L}{2}\right)\sigma_3
\eeq
where $\sigma_3$ is the third Pauli matrix and the time dependence is implicit.  This boundary condition is easily visualized using the Hopf coordinates
\bea
U&=&\left(\begin{tabular}{cc}$z_1$&$iz_2$\\$i\overline{z_2}$&$\overline{z_1}$
\end{tabular}\right)\hsp z_1=\cos(\theta)e^{i\phi_1}\nonumber\\z_2&=&\sin(\theta)e^{i\phi_2},\ 
\theta\in[0,\pi/2],\ \phi_i\in [0,2\pi] \label{u}\nonumber
\eea
where it is just
\beq
\phi_2\left(\frac{L}{2}\right)=\phi_2\left(-\frac{L}{2}\right)+\pi. \label{bordo}
\eeq
The boundary condition is trivial when $U$ commutes with $\sigma_3$, corresponding to the circle
\beq
U={\mathrm{exp}}\left(i\phi_1\sigma_3\right) \label{circ}
\eeq
or equivalently to the circle $(\theta,\phi_1)=(0,\phi_1)$, where the $\phi_2$ circle degenerates.

As described in Ref.~\cite{daniele}, the twisted boundary conditions increase the energy of a configuration away from these fixed points, and so lead to a potential for $\theta$.  Classically this potential vanishes precisely at the fixed point set of the symmetry $\phi_2\rightarrow\phi_2+\pi$, and so the circle (\ref{circ}) is the classical vacuum manifold of this theory.  It is connected.  

What about the minima of the effective potential, obtained by integrating out the oscillations transverse to this vacuum manifold?  In principle $\phi_1$-dependent masses for these transverse oscillations could lead to a $\phi_1$-dependent effective potential.  This new potential could lift part of the circle, leaving a space of minima which is no longer connected.  However in the case at hand, both the action and also the boundary condition are invariant under shifts $\phi_1\rightarrow\phi_1+c$.  As we are in 2-dimensions, spontaneous symmetry breaking is forbidden \cite{coleman} and so this symmetry is also respected by the effective action.  Therefore the minima of the effective potential will have the same symmetry.  In the weakly coupled limit, the minima of the effective potential must be a nontrivial subset of the classical vacuum manifold but the only such subset preserving the shift symmetry is the entire circle.  Therefore the space of minima of the effective action is a circle, which is connected and so it does not satisfy the criterion described the abstract for a nonperturbative mass gap.  

This is not to exclude nonperturbative contributions to the mass gap.  Indeed, such contributions are expected.  However, as the space of minima is compact we expect perturbative contributions to the mass gap.  As this theory is weakly coupled, the perturbative contributions will be far larger than the nonperturbative contributions, and so we say that the mass gap is perturbatively generated.  Below we will calculate these perturbative contributions explicitly and see that they are nonvanishing.

Ref.~\cite{daniele} uses the Hopf coordinates with the fundamental domain $\theta\in[0,\pi]$,\ $\phi_1\in [0,\pi]$,\ $\phi_2\in [0,2\pi]$.  In these coordinates the boundary condition is still given by Eq.~(\ref{bordo}).  However now the fixed point set is $\sin(\theta)=0$ where $\phi_2$ degenerates.  In terms of $\theta$ and $\phi_1$ this consists of two intervals $(\theta=0,\phi_1\in[0,\pi])$ and $(\theta=\pi,\phi_1\in[0,\pi])$.  It was claimed that there are two near degenate vacua which are supported on these two intervals with even and odd parity under the symmetry $\theta\mapsto\pi-\theta$.  However the points $(\theta,\phi_1)=(0,\pi)$ and $(\theta,\phi_1)=(\pi,0)$ both correspond to the same point $(z_1,z_2)=(-1,0)$ while both $(\theta,\phi)=(0,0)$ and $(\theta,\phi_1)=(\pi,\pi)$ correspond to the same point $(z_1,z_2)=(1,0)$ therefore these two intervals are connected at their endpoints.  The union of these two intervals is a circle, indeed it is just the vacuum manifold found using the fundamental domain in Eq.~(\ref{u}).  The excitations of fields on this circle correspond to the states of a particle in a periodic box.  In particular a state which is odd under $\theta\mapsto\pi-\theta$, or equivalently $\phi_1\mapsto\phi_1+\pi$, will correspond to an odd excitation of the particle in a box, while the ground state is an even function.  This splitting is {\it{perturbative}}, and in fact requires no deep excursions into the classically forbidden zone in which $\sin(\theta)>0$.

As was shown in Ref.~\cite{daniele}, at small $L$ this theory is weakly coupled and the probability for the particle to venture far from the fixed point is exponentially surpressed.  The interactions correspond to the curvature of the geometry and so the weak coupling limit corresponds to a flattened neighborhood of the fixed circle.  More precisely, in the small $L$ limit the target space becomes $\C\times S^1$ where $z_2$ is a coordinate of the $\C$ and $\phi_1$ is a coordinate of the $S^1$.  The $\C$ and $S^1$ sectors are decoupled from each other at weak coupling.  The twisted boundary conditions only affect the $\C$, where they yield $z_2(L/2)=-z_2(-L/2)$.  

Expanding $z_2=y_1+iy_2$, the boundary condition becomes $y_i(L/2)=-y_i(-L/2)$.  From the action
\bea
S&=&\frac{1}{2g^2}\int dx dt {\bf{Tr}}\left(\partial_\mu U^\dagger \partial^\mu U\right)\\
&=&\frac{1}{g^2}\int dx dt(\partial_\mu\phi_1\partial^\mu\phi_1+\sum_i \partial_\mu y_i\partial^\mu y_i)\nonumber
\eea
one can find the canonical momenta
\beq
\pi=\frac{2}{g^2}{\partial_t\phi_1}\hsp
\Pi_i=\frac{2}{g^2}{\partial_t y_i}.
\eeq
The quantization of $\phi_1$ is just that of a particle in a periodic box.  Suppressing time dependence, $\phi_1$ can be Fourier expanded on the compactified circle $x$
\bea
\phi_1=\phi_1^{(0)}+\sum_{n\neq 0}\frac{1}{\sqrt{2 \frac{2\pi }{L}n}}\left(a_n+a^\dagger_{-n}\right) e^{i\frac{2\pi  x}{L}n}\nonumber\\
\pi=\pi^{(0)}-\frac{2i}{g^2}\sum_{n\neq 0}\sqrt{\frac{2\pi n}{2L}}\left(a_n-a^\dagger_{-n}\right) e^{i\frac{2\pi x}{L}n}
\eea
Imposing $[\phi_1(x_1),\pi(x_2)]=i\delta(x_1-x_2)$ yields the commutation relations
\beq
[\phi_1^{(0)},\pi^{(0)}]=\frac{i}{L}\hsp
[a_m,a^\dagger_n]=\frac{g^2}{2L}\delta_{mn}.
\eeq

Normal ordering the Legendre transform one obtains the Hamiltonian
\beq
H=\frac{g^2L}{4}\pi^{(0)}\pi^{(0)}+\frac{4\pi}{g^2}\sum_{n\neq 0} |n|a^\dagger_n a_n.
\eeq
Let the vacuum state be annihilated by both $a_n$ and $\pi^{(0)}$.  Then there will be two families of raising operators which create excited states.  First $e^{in\phi_1^{(0)}}$ is well-defined for $n$ an integer as $\phi_1$ is $2\pi$-periodic.  These are the excited oscillator states of Ref.~\cite{daniele} and, in agreement with Eq.~(5.18), their energy is
\beq
[H,e^{in\phi_1^{(0)}}]=E_n e^{in\phi_1^{(0)}}\hsp E_n=\frac{g^2n^2}{4L}
\eeq
which is the perturbative result that one expects for a particle in a box.  Note that the lowest level state which is odd under $\phi_1\mapsto\phi_1+\pi$ is the state $n=1$, yielding a mass gap of $g^2/4L$.  This is our main result: the mass gap is perturbative.

The Kaluza-Klein (KK) modes also yield excited states, created by $a^\dagger_n$.  Their energy is given by
\beq
[H,a_n^\dagger]=E^\prime_n a_n^\dagger\hsp E^\prime_n=2\pi\frac{n}{L}.
\eeq
Note that $E^\prime$ is $g$-independent, unlike $E$, and so in the small $g$ or equivalently the small $L$ limit, these KK modes are much heavier than the particle in a box excitations.

The antiperiodic boundary conditions on the fields $y_i$ yield the Fourier decomposition
\bea
y_i&=&\\
&&\hspace{-1.0cm}\sum_{n}\frac{1}{\sqrt{2 \frac{2\pi}{L}(n+\frac{1}{2})}}\left(b_{i,n+\frac{1}{2}}+b^\dagger_{i,-n-\frac{1}{2}}\right) e^{i\frac{2\pi x}{L}(n+\frac{1}{2})}\nonumber\\
\Pi_i&=&\nonumber\\
&&\hspace{-1.0cm}-\frac{2i}{g^2}\sum_{n}\sqrt{\frac{2\pi }{2L}(n+\frac{1}{2})}\left(b_{i,n+\frac{1}{2}}-b^\dagger_{i,-n-\frac{1}{2}}\right) e^{i\frac{2\pi  x}{L}(n+\frac{1}{2})}.\nonumber
\eea
Again the commutation relations of the quantum mechanical modes follow from those of the quantum fields
\bea
[y_i(x_1),\Pi_j(x_2)]&=&i\delta_{ij}\delta(x_1-x_2)\nonumber\\
\hspace{0cm}
[b_{i,m+\frac{1}{2}},b^\dagger_{j,n+\frac{1}{2}}]&=&\delta_{ij}\delta_{mn}\frac{g^2}{2L}.
\eea
One then finds the Hamiltonian as above
\bea
H&=&\int dx\sum_i(\frac{g^2}{4}:\Pi_i\Pi_i:+\frac{1}{g^2}:\partial_x y_i\partial_x y_i:)\nonumber\\
&=&\frac{4\pi}{g^2}\sum_{i,n}\left|n+\frac{1}{2}\right|b^\dagger_{i,n+\frac{1}{2}}b_{i,n+\frac{1}{2}}.
\eea
Excitated states are created with $b^\dagger_{i,n+\frac{1}{2}}$ each of which increases the energy by $\overline{E}_n$
\beq
[H,b^\dagger_{i,n+\frac{1}{2}}]=E_nb^\dagger_{i,n+\frac{1}{2}}\hsp
E_n=\frac{4\pi}{L}\left(n+\frac{1}{2}\right).
\eeq

Now we turn to the $\cp^1$ model.  Note that the $y_i$ alone also describe the weak coupling limit of the $\cp^1$ sigma model with antiperiodic boundary conditions introduced in Ref.~\cite{cplungo,cpcorto}.   As $\cp^1$ is $S^2$ and SU(2) is an $S^3$, one can pass from one model to the other via the Hopf projection $S^3\rightarrow S^2$ which identifies $(\phi_1,\phi_2)\sim(\phi_1+\alpha,\phi_2+\alpha)$.  The invariant angle $\phi=\phi_1-\phi_2$ is the azymuthal coordinate of the $S^2$ and as such it degenerates at the poles $\theta=0$ and $\theta=\pi/2$.  The twisted boundary conditions are $\phi(L/2)=\phi(-L/2)+\pi$ and so are trivial at the two poles, which are the classical vacua of the theory.  At weak coupling or more precisely small $L$, each of these classical vacua is described by the $y_i$ theory described above.

We can describe these two weak-coupling vacua explicitly by decomposing the field $y_i$ into KK modes, the degrees of freedom in the corresponding quantum mechanics,
\bea
y_i=\sum_n y_{i,n+\frac{1}{2}}e^{i\frac{2\pi x}{L}\left(n+\frac{1}{2}\right)}\nonumber\\
\Pi_i=\sum_n \Pi_{i,n+\frac{1}{2}}e^{i\frac{2\pi x}{L}\left(n+\frac{1}{2}\right)}
\eea
whose commutation relations yield a simple Schr\"odinger representation
\bea
[y_{i,m+\frac{1}{2}},\Pi_{j,n+\frac{1}{2}}]&=&\delta_{ij}\delta_{m,-n}\frac{i}{L}\nonumber\\
\Pi_{i,n+\frac{1}{2}}&=&-\frac{i}{L}\frac{\partial}{\partial y_{-n-\frac{1}{2}}}.
\eea
The vacuum must be annihilated by all of the $b$'s
\bea
0&=&b_{i,n+\frac{1}{2}}|0\rangle\\&\propto& \left[2\pi g^2\left(n+\frac{1}{2}\right)y_{i,n+\frac{1}{2}}+\frac{\partial}{\partial y_{i,-n-\frac{1}{2}}}\right]|0\rangle\nonumber
\eea
and so it is proportional to
\beq
\psi={\mathrm{exp}}\left[-\frac{2\pi}{g^2}\sum_{i,n}\left|n+\frac{1}{2}\right| |y_{i,n+\frac{1}{2}}|^2
\right] \label{onda}
\eeq
where we have used $y_{i,n+\frac{1}{2}}=y^*_{i,-n-\frac{1}{2}}$ which is a consequence of the reality of $y_i$.  Eq.~(\ref{onda}) may be interpreted as a wave function of an infinite dimensional quantum mechanics or equivalently \cite{stuck,friedrichs} as the Schr\"odinger wave functional of the compactified quantum field theory.  One may observe that, as expected from a product harmonic oscillators, states are exponentially confined to the classical vacuum with higher KK modes $n$ more strongly confined.   In general the distance that states may wander from the vacuum is of order $g$.

The lightest modes are $n=-1$ and $n=0$ which are related by complex conjugation.  Although this free truncation experiences corrections (to the exponential) of order unity far from the vacuum, one may crudely estimate the overlap of the two vacua by inserting $y\sim\pi/2$ to conclude that indeed the overlap is of order exp$(-c/g^2)$ for some $c$, as expected from a [fractional] instanton effect.

The generalization to a nonlinear sigma model with target space metric $g_{ij}$ is straightforward.  In this case
\bea
\Pi_i&=&\frac{2}{g^2}g_{ij}{\partial_t y_j}\\
\mathcal{H}&=&\sum_{i,j}(\frac{g^2}{4}:g^{ij}\Pi_i\Pi_j:+\frac{1}{g^2}:g_{ij}\partial_x y_i\partial_x y_j:) \nonumber\label{geqs}
\eea
where $g^{ij}$ is the inverse metric.  In the case of a $\cp^1$ model, we identify $y_1+iy_2$ with the affine coordinates for $\cp^1$.  Now one classical vacuum is at the origin while the other lies at infinity.  As the $\cp^1$ is a unit sphere, in affine coordinates the metric is given by four times the Fubini study metric
\beq
g_{ij}=\frac{4\delta_{ij}}{(1+y_1^2+y_2^2)^2}.
\eeq

Let us now truncate our theory down to the four lowest KK modes, corresponding to $|n+1/2|=1/2$.  Note that this truncation explicitly violates the $y_1+iy_2\rightarrow 1/(y_1+i y_2)$ symmetry which exchanges the vacua. Now our two quantum fields reduce to four-dimensional quantum mechanics via the decomposition
\beq
y_i=\sqrt{\frac{L}{2\pi}}\left[\left(b_{i,-\frac{1}{2}}+b^\dagger_{i,\frac{1}{2}}\right)e^{-i\frac{\pi}{L}x}+\left(b_{i,\frac{1}{2}}+b^\dagger_{i,-\frac{1}{2}}\right)e^{i\frac{\pi}{L}x}\right].
\eeq
This 4-dimensional theory is invariant under rotations of $\phi$ or equivalently $y_1+iy_2$.   The low lying states will be rotation-invariant and these are already sufficient to study the instantons.  Therefore we will fix the rotational freedom by setting $b_{1,1/2}=-b_{1,-1/2}$ so that $y_{1}$ is imaginary and equal to
\beq
y_1=-2i\sqrt{\frac{L}{2\pi}}\left(b_{1,\frac{1}{2}}-b^\dagger_{1,\frac{1}{2}}\right)\mathrm{sin}\left(\frac{\pi}{L}x\right).
\eeq
Physically, this means that the state reaches its maximal extent in $y_1$ at $|x|=L/2$.  By combining a rotation with a shift in $x$ we can also impose the condition $b_{2,1/2}=b_{2,-1/2}$ so that $y_{2,1/2}$ is real.  This corresponds to an orbit in which $y_1$ and $y_2$ are the principle axes, with $y_2$ extremized at $x=0$ and vanishing at the boundaries.
We are left with a 2-dimensional quantum mechanics in which the field $y_2$ has been decomposed as
\beq
y_2=2\sqrt{\frac{L}{2\pi}}\left(b_{2,\frac{1}{2}}+b^\dagger_{2,\frac{1}{2}}\right)\mathrm{cos}\left(\frac{\pi}{L}x\right).
\eeq
Now that the mode numbers are all equal to $1/2$, they will be omitted.  The conjugate momenta may be decomposed
\bea
\Pi_i&=&\frac{2}{g^2}\sqrt{\frac{2\pi}{L}}\\
&&\hspace{-1cm}\times \left[g_{i1} \left(b_1+b^\dagger_1\right){\mathrm{sin}}\left(\frac{\pi}{L}x\right)-ig_{i2}\left(b_2-b^\dagger_2\right){\mathrm{cos}}\left(\frac{\pi}{L}x\right)\right]\nonumber
\eea

Putting everything together we obtain the Hamiltonian
\begin{widetext}
\bea
\mathcal{H}&=&\frac{8\pi}{g^2L}\frac{\left[\left(b_1+b^\dagger_1\right)^2+\left(b_2+b^\dagger_2\right)^2\right]{\mathrm{sin}}^2\left(\frac{\pi}{L}x\right)
-\left[\left(b_1-b^\dagger_1\right)^2+\left(b_2-b^\dagger_2\right)^2\right]{\mathrm{cos}}^2\left(\frac{\pi}{L}x\right)}
{\left(1+4\frac{L}{2\pi}\left[-\left(b_1-b^\dagger_1\right)^2{\mathrm{sin}}^2\left(\frac{\pi}{L}x\right)+\left(b_2+b^\dagger_2\right)^2{\mathrm{cos}}^2\left(\frac{\pi}{L}x\right)
\right]\right)^2}.\nonumber\\
H&=&\int_{x=-L/2}^{L/2}dx \mathcal{H}=\frac{4\pi}{L}\frac{1}{\sqrt{\left(1+b_{2+}^2\right)\left(1+b_{1-}^2\right)}}
\left[\frac{b_{1+}^2+b_{2+}^2}{1+b_{1-}^2}+\frac{b_{1-}^2+b_{2-}^2}{1+b_{2+}^2}\right] \label{qmh}
\eea
\end{widetext}
where we have defined
\beq
b_{i+}=\frac{\sqrt{L}}{g}(b_i+b_i^\dagger)\hsp
b_{i-}=-i\frac{\sqrt{L}}{g}(b_i-b_i^\dagger)
\eeq
Note that (\ref{qmh}) has a simple interpretation as a Hamiltonian for 2-dimensional quantum mechanics with coordinates  $b_{1-}$ and $b_{2+}$ and momenta $-b_{1+}$ and $b_{2-}$.  The isolated vacua are at $b_{1-}=b_{2+}=0$ and $b_{1-}=b_{2+}=\infty$.  Using this truncated Hamiltonian, one may calculate the instanton contributions to the wave function and energies.

Unfortunately $[b_{i+},b_{j-}]=i\delta_{ij}$ only near the vacuum at the origin $y=0$ and so in general these positions and momenta are not quite canonically conjugate.  This is a result of the metric in the expression for $\Pi_i$ in Eq.~(\ref{geqs}), which differs from the identity matrix away from the origin. 

In general the dynamics of this theory is quite complicated.  The mode expansion truncation does not commute with the QFT Hamiltonian, although the difference is subleading in $g$, and so the dynamics of the truncated QM and the original QFT are generally inequivalent.  One exception is the trajectories $b_{1-}=b_{2+}$, representing maps where the latitude is independent of $x$.  Such trajectories interpolate between the vacua at infinity and zero.  The half-charged instanton is of this form in the Euclidean theory.
                                   
Beyond the leading order interactions, the Hamiltonian (\ref{qmh}) differs from that found in Refs.~\cite{danielecorto,daniele,cplungo,cpcorto}.  Note that Eq.~(3.17) of Ref.~\cite{cplungo} is not consistent with the condition that the field be restricted to the $\cp^1$, since the field $\tilde{n}$ in that equation is not in general a unit vector.   This can be corrected by adding a constraint by hand to the Langrangian \cite{bardeen76}, or by introducing a Langrange multiplier \cite{divecchia78}  or Dirac constraints \cite{lapa}. More importantly, as is explained under Eq.~(4.20) of \cite{cplungo}, in the reduction to quantum mechanics it is assumed that the latitude $\theta$ is constant. This implies that fixed time slices are circles of parallel in $\cp^1$, not geodesics, and so in general not the lowest energy curves with given boundaries.  In other words, they only consider configurations with $b_{1-}=b_{2+}$, yielding a one-dimensional slice of the quantum mechanical system which includes the half-instantons of interest but not the lowest energy perturbative excitations.

\section* {Acknowledgement}

\noindent
JE is supported by the CAS Key Research Program of Frontier Sciences grant QYZDY-SSW-SLH006 and the NSFC MianShang grants 11875296 and 11675223.  JE also thanks the Recruitment Program of High-end Foreign Experts for support.


\end{document}